# Sudden synchrony leaps accompanied by frequency multiplications in neuronal activity

Roni Vardi[1*], Amir Goldental[2*], Shoshana Guberman[1,2*], Alexander Kalmanovich[1], Hagar Marmari[1] and Ido Kanter[1,2]

[1]Gonda Interdisciplinary Brain Research Center, and the Goodman Faculty of Life Sciences, Bar-Ilan University, Ramat-Gan 52900, Israel. [2]Department of Physics, Bar-Ilan University, Ramat-Gan 52900, Israel.

*R.V. A.G. and S.G. contributed equally to this work.

Correspondence should be addressed to either of the following: Roni Vardi, E-mail: ronivardi@gmail.com; Ido Kanter, E-mail: ido.kanter@biu.ac.il

A classical view of neural coding relies on temporal firing synchrony among functional groups of neurons; however the underlying mechanism remains an enigma. Here we experimentally demonstrate a mechanism where time-lags among neuronal spiking leap from several tens of milliseconds to nearly zero-lag synchrony. It also allows sudden leaps out of synchrony, hence forming short epochs of synchrony. Our results are based on an experimental procedure where conditioned stimulations were enforced on circuits of neurons embedded within a large-scale network of cortical cells in vitro and are corroborated by simulations of neuronal populations. The underlying biological mechanisms are the unavoidable increase of the neuronal response latency to ongoing stimulations and temporal or spatial summation required to generate evoked spikes. These sudden leaps in and out of synchrony may be accompanied by multiplications of the neuronal firing frequency, hence offering reliable information-bearing indicators which may bridge between the two principal neuronal coding paradigms.

## INTRODUCTION

One of the major challenges of modern neuroscience is to elucidate the brain mechanisms that underlie firing synchrony among neurons. Such spike correlations with differing degrees of temporal precision have been observed in various sensory cortical areas, in particular in the visual (Eckhorn et al., 1988; Gray et al., 1989), auditory (Ahissar et al., 1992; Nicolelis et al., 1995), somatosensory (Nicolelis et al., 1995) and frontal (Vaadia et al., 1995) areas. Several mechanisms have been suggested, including the slow and limited increase in neuronal response latency per evoked spike (Vardi et al., 2013b). On a neuronal circuit level its accumulative effect serves as a non-uniform gradual stretching of the effective neuronal circuit delay loops.

Consequently, small mismatches of only a few milliseconds among firing times of neurons can vanish in a very slow gradual process consisting of hundreds of evoked spikes per neuron.

The phenomenon of sudden leaps from firing mismatches of several tens of milliseconds to nearly zero-lag synchronization, below a millisecond, is counterintuitive. Since the dynamical variations in neuronal features, e.g. the increase in neuronal response latencies per evoked spike, are extremely small, one might expect only very slow variations in firing timings. Moreover, relative changes among firing times of neurons require dynamic relaxation of the entire neuronal circuit to achieve synchronization. Hence, sudden leaps, in and out of synchrony, seem unexpected.

In the present study, we propose a new experimentally corroborated mechanism allowing leaps in and out of synchrony. The procedure is based on conditioned stimulations enforced on neuronal circuits embedded within a large-scale network of cortical cells in vitro (Marom and Shahaf, 2002; Morin et al., 2005; Wagenaar et al., 2006; Vardi et al., 2012). These stimulations varied in strength, so that the evoked spikes of selected neurons required temporal summation. We demonstrate that the underlying biological mechanism to sudden leaps in and out of synchrony is the unavoidable increase of the neuronal response latency (Aston-Jones et al., 1980; De Col et al., 2008; Ballo and Bucher, 2009; Gal et al., 2010) to ongoing stimulations, which imposes a non-uniform stretching of the neuronal circuit delay loops.

## MATERIAL AND METHODS

### CULTURE PREPARATION

Cortical neurons were obtained from newborn rats (Sprague-Dawley) within 48 h after birth using mechanical and enzymatic procedures (Marom and Shahaf, 2002; Vardi et al., 2012; Vardi et al., 2013b). All procedures were in accordance with the National Institutes of Health Guide for the Care and Use of Laboratory Animals and Bar-Ilan University Guidelines for the Use and Care of Laboratory Animals in Research and were approved and supervised by the Institutional Animal Care and Use Committee.

The cortex tissue was digested enzymatically with 0.05% trypsin solution in phosphate-buffered saline (Dulbecco's PBS) free of calcium and magnesium, supplemented with 20 mM glucose, at 37°C. Enzyme treatment was terminated using heat-inactivated horse serum, and cells were then mechanically dissociated. The neurons were plated directly onto substrate-integrated multi-electrode arrays (MEAs) and allowed to develop functionally and structurally mature networks over a time period of 2-3 weeks in vitro, prior to the experiments. Variability in the number of cultured days in this range had no effect on the observed results. The number of plated neurons in a typical network is in the order of 1,300,000, covering an area of about 380 mm$^2$. The preparations were bathed in minimal essential medium (MEM-Earle, Earle's Salt Base without L-Glutamine) supplemented with heat-inactivated horse serum (5%), glutamine (0.5 mM), glucose (20 mM), and gentamicin (10 g/ml), and maintained in an atmosphere of 37°C, 5% $CO_2$ and 95% air in an incubator as well as during the electrophysiological measurements. All experiments were conducted on cultured cortical neurons that were functionally isolated from their network by a pharmacological block of glutamatergic and GABAergic synapses. For each plate, 12-20 μl of a cocktail of synaptic blockers was used, consisting of 10 μM CNQX (6-cyano-7-nitroquinoxaline-2,3-dione), 80 μM APV (amino-5-phosphonovaleric acid) and 5 μM Bicuculline. This cocktail did not block the spontaneous network activity completely, but rather made it sparse. At least one hour was allowed for stabilization of the effect.

**MEASUREMENTS AND STIMULATION**

An array of 60 Ti/Au/TiN extracellular electrodes, 30 μm in diameter and spaced either 200 or 500 μm from each other (Multi-ChannelSystems, Reutlingen, Germany) was used. The insulation layer (silicon nitride) was pre-treated with polyethyleneimine (Sigma, 0.01% in 0.1 M Borate buffer solution). A commercial setup (MEA2100-2x60-headstage, MEA2100-interface board, MCS, Reutlingen, Germany) for recording and analyzing data from two 60-electrode MEAs was used, with integrated data acquisition from 120 MEA electrodes and 8 additional analog channels, integrated filter amplifier and 6-channel current or voltage stimulus generator (for both MEAs). Mono-phasic square voltage pulses (-900 – -100 mV, 100-500 μs) were applied through extracellular electrodes. Each channel was sampled at a frequency of 50k sample/s. Action potentials were detected on-line by threshold crossing. For each of the recording channels a threshold for spike detection was defined separately, prior to the beginning of the experiment.

**CELL SELECTION**

Each circuit node was represented by a stimulation source (source electrode) and a target for the stimulation – the recording electrode (target electrode). These electrodes (source and target) were selected as the ones that evoked well-isolated, well-formed spikes and reliable response with high signal-to-noise ratio. This examination was done with stimulus intensity of -800 mV using 30 repetitions at a rate of 5Hz followed by 1200 repetitions at a rate of 10Hz.

**STIMULATION CONTROL**

A node response was defined as a spike occurring within a typical time window of 2-10 ms following the electrical stimulation. The activity of all source and target electrodes was collected, and entailed stimuli were delivered in accordance to the circuit connectivity.

**Circuit connectivity, $\tau$:** Conditioned stimulations were enforced on the circuit neurons embedded within a large-scale network of cortical cells in vitro, according to the circuit connectivity. Initially, each delay was defined as the expected time between the evoked spikes of two linked neurons; e.g. conditioned to a spike recorded in the target electrode assigned to neuron A, a spike will be detected in the target electrode of neuron B after $\tau_{AB}$ ms. For this end, conditioned to a spike recorded in the target electrode of neuron A, a stimulus will be applied after ($\tau_{AB}$-$L_B(0)$) ms to the source electrode of neuron B, where $L_B(0)$ is the initial latency of neuron B.

In cases where missed evoked spikes caused a termination of the neuronal circuit activity, stimulation was given to neuron A after a period of 100 ms, to restart the circuit's activity.

All neurons were stimulated at a rate of 10 Hz (**Figure 1** and **Figure 3**) or 8 Hz (**Figure 2**), before the leap to synchronization.

Strong stimulations, (-800 mV, 200 μs), resulting in a reliable neural response, were given to all circuit neurons excluding neuron C (**Figure 1** and **Figure 2**) and E (**Figure 3**). Weak stimulations (**Figure 1**: -450 mV, 40 μs. **Figure 2**: -600 mV, 60 μs. **Figure 3**: -700 mV, 60 μs) were given to neuron C (**Figure 1** and **Figure 2**) or E (**Figure 3**), so that an evoked spike is expected only if the time-lag between two consecutive weak stimulations is short enough. In cases where the time-lag between two consecutive stimulations was shorter than 20 μs (from the end of the first stimulation to the beginning of the consecutive one), a unified strong stimulation was applied, to overcome technical limitations. The weak stimulations were defined for each neuron separately, due to differences in their threshold.

$T_{TS}$ (TS stands for temporal summation) is the maximal time-lag between two weak stimulations which typically results in an evoked spike. This quantity was empirically estimated by gradually changing the time-lag between two weak stimulations, and found to differ between neurons.

## DATA ANALYSIS

Analyses were performed in a Matlab environment (MathWorks, Natwick, MA, USA). Action potentials were detected by threshold crossing. In the context of this study, no significant difference was observed in the results under threshold crossing or voltage minima for spike detection. Reported results were confirmed based on at least ten experiments each, using different sets of neurons and several tissue cultures.

## RESULTS

### LEAP TO SYNCHRONY ACCOMPANIED BY A DOUBLED FIRING FREQUENCY

#### EXPERIMENTAL RESULTS

We first demonstrate leaps to synchrony using a neuronal circuit consisting of four neurons and conditioned stimulations split into weak/strong stimulations (**Figure 1A**). A strong stimulation consists of a relatively high amplitude and/or relatively long pulse duration such that an evoked spike is generated reliably, whereas a weak stimulation consists of a lower amplitude and/or pulse duration, such that an evoked spike is expected only if the time-lag between two consecutive weak stimulations is short enough. All delays (denoted on connecting lines between neurons in **Figure 1A**) were selected to initially include the response latency of the target neuron, e.g. the time-lag from neuron A to B, $\tau_{AB}$, was initially set to $\tau$-$L_B(0)$ where $L_B(0)$ stands for the initial response latency of neuron B. For $\tau=50$ ms, neurons A and B initially fire alternately, in and out of phase, at a frequency of ~10 Hz (**Figure 1B**). Neuron D fires ~$\tau/2$ ms laggard to neuron A (**Figure 1C**) and the time-gap between two weak stimulations arriving at neuron C, $\Delta(Stim_C)$, is initially $\varepsilon$ (**Figures 1A,B**). The experimentally estimated maximal time-gap between stimulations of neuron C which generates an evoked spike (temporal summation) is denoted by $T_{TS}$, thus for $\Delta(Stim_C) > T_{TS} \approx 0.23$ ms neuron C typically does not fire. As a result of the increase in the response latency of neuron D, $\Delta(Stim_C)$ is reduced (green-line **Figure 1B**) sufficiently so that neuron C starts firing ($\Delta(Stim_C) \leq T_{TS}$) (**Figure 1C**). The circuit now consists of two delay loops, ~$2\tau$ (A-B-A) and ~$3\tau$ (A-C-B-A). Since the greatest common divisor (GCD) of the circuit delay loops is GCD(2,3)=1, conditioned to the firing of neuron C, zero-lag synchronization between neurons A and B is theoretically expected (Kanter et al., 2011) after a very short transient, $\tau$ (**Figure 1C**). This phenomenon is clearly demonstrated by the leap in the time-lag between the spikes of neurons A and B, $Sync_{AB}$ (blue line in **Figure 1B**), and is accompanied by a sudden frequency multiplication from ~10 Hz to ~20 Hz (**Figure 1C**). The sudden emergence of $Sync_{AB} \approx 0$ ms requires only a single firing of neuron C, and is then maintained by the mutual firing of neurons A and B, independently of the firing of neuron C (**Figure 1C**). For a given $T_{TS}$, the number of evoked spikes of neuron D until the leap to synchrony, n, increases with $\varepsilon$ (**Figure 1D**). Quantitatively, using the experimental response latency profile of neuron D, $L_D$, one can find n fulfilling the equality:

$$\Delta L_D(n) \approx \varepsilon - T_{TS} \qquad (1)$$

where $\Delta L_D(n)$ stands for the increase in response latency of neuron D after n evoked spikes (**Figure 1E**). Note that neuron D is laggard to neuron A, thus the number of evoked spikes of neuron A until the leap to synchrony increases with $\varepsilon$ as well, in accordance with Equation 1 (**Figure 1D**). Since $T_{TS}$ varies between neurons and even within the same neuron over different trials, deviations from this equation are expected (e.g. $\Delta L_D$ for $\varepsilon=0.8$ ms and $\varepsilon=1$ ms are almost the same, **Figures 1D,E**). A slow gradual increase in $Sync_{AB}$ after a leap to synchrony (**Figure 1D**) is theoretically attributed to the difference in the increase of neuronal response latencies $|\Delta L_A(n)-\Delta L_B(n)|$ and the leap out of synchrony (**Figure 1D**) is a consequence of a response failure of neurons A and/or B (see Section "Slow Divergence out of Synchrony" in Appendix). Similar results were obtained and exemplified for spatial summation (not shown), where weak stimulations were given to a neuron *through two different source electrodes*. An evoked spike is expected only if the time-lag between two consecutive weak stimulations, controlled by the relative stimulation timings of the source electrodes, is short enough. Note that in order to identify sudden leaps in or out of synchrony, as well as the effect of a single neuronal response failure on synchronization, statistical measures of synchrony (e.g. Kreuz et al., 2007; Shimokawa and Shinomoto, 2009) are insufficient.

**SIMULATIONS OF POPULATION DYNAMICS**

The sudden leap to synchrony was experimentally verified under the limitation where each circuit node is represented by a single neuron, and is demonstrated to be robust under simulations of population dynamics (**Figures 1F,G**). Each one of the four nodes (**Figure 1A**) now represents a population comprised of 40 Hodgkin-Huxley sparsely connected neurons (for simulation details, see Vardi et al., 2013a). For the parameters used, $T_{TS} \approx 1.3$ ms, $\varepsilon=2$ ms and 0.2 ms variance for the Gaussian distribution of the delays, a leap to synchrony is expected following Equation 1 after ~20 spikes of cluster A (**Figure 1F**). The simulated $Sync_{AB}$ is defined as the absolute difference between the average spiking times of the neurons comprising clusters A and B, where at least 50% of the neurons in a cluster fired (**Figure 1G**). Initially, several neurons in cluster C fire as a result of relatively close stimulations from either cluster A or D. This sporadic firing is a consequence of the Gaussian distribution of the delays between populations; however, their impact on the firing activity of cluster B is negligible. As neurons of cluster D fire repeatedly, $\Delta(Stim_C)$ decreases and more neurons from cluster C fire. Consequently, the activity of cluster C is enhanced such that a leap to synchrony is observed, accompanied by frequency doubling from ~10Hz to ~20Hz (**Figures 1F,G**). A leap out of synchrony was not observed in the simulations, since population dynamics are more robust to a single neuron's response failure in comparison to a neuronal circuit where each node is represented by a single neuron (**Figure 1A,D**). Low connectivity, as well as a wider Gaussian distribution of delays between populations are expected to enhance fluctuations and response failures, and will eventually lead to a leap out of synchrony.

Population dynamics exhibit consistency with most of the experimental results, hence minimizing the possibility of these results as being only an artifact of the tissue culture. Nevertheless, the verification of our results in more realistic scenarios is required, including shorter delays and their interplay with the neuronal refractory period, the morphology of the neurons instead of considering neurons as points (Doiron et al., 2006), as well as possible adaptation mechanisms in the form of short and long term synaptic plasticity (Abbott and Regehr, 2004; Izhikevich, 2006).

**LEAP TO SYNCHRONY ACCOMPANIED BY TRIPLED FIRING FREQUENCY**

More general features of a sudden leap to synchrony are exemplified by increasing the delay from neuron B to A, $\tau_{BA}$, from $\tau$ (**Figure 1A**) to $2\tau$ (**Figure 2A**). The circuit now consists of two delay loops, $\sim 3\tau$ (A-B-A) and $\sim 4\tau$ (A-C-B-A) (**Figure 2C**). Since GCD(4,3)=1, zero-lag synchronization is theoretically expected, conditioned to the firing of neuron C. Initially, Neurons A and B fire at a frequency of $\sim$8 Hz ($3\tau$=125 ms) (**Figure 2C**) and $Sync_{AB} \approx \tau$ (**Figure 2B**). Neuron C starts to fire as $\Delta(Stim_C) \leq T_{TS} \approx 0.2$ ms, resulting in $Sync_{AB} \approx 0$ which is accompanied by tripled firing frequency (**Figure 2C**). The number of evoked spikes by neuron D (or its leader neuron A) to the leap increases with $\varepsilon$ in a nonlinear manner following $\Delta L_D(n)$, in accordance with Equation 1 (**Figures 2D,E**).

Typically, several leaps in and out of synchrony between neurons A and B occur before arriving at a stable nearly zero-lag synchronization (**Figure 2D**). These oscillations are attributed to unreliable responses of neuron C, and increase the duration of the relaxation to synchrony (**Figure 2D**). Similar oscillations on the way out of synchrony (**Figure 2D**) are attributed to the first response failure of either neuron A or B. Consequently, neurons A and B fire alternately in time-lags $\tau$ and $2\tau$. The final exit out of synchrony occurs in the second response failure of neurons A or B.

Simulation results (**Figures 2F,G**) confirmed the robustness of the experimentally observed leap to synchrony in population dynamics. The oscillations in the relaxation to synchrony are attributed to response failures of cluster C. These failures are a consequence of fluctuations in the firing timings of clusters A and D and the Gaussian distribution of their delays to cluster C.

**EPOCHS OF SYNCHRONY NOT ACCOMPANIED BY A CHANGE IN FREQUENCY**

A mechanism to leap out of synchrony as well as the interrelation between the sudden leap to synchrony and the firing frequency are at the center of the next examined neuronal circuit (**Figure 3A**). This circuit consists solely of a $2\tau$-delay loop, hence neurons A and F fire alternately in $\sim\tau$ ms time-lags. Nevertheless, neuron A affects neuron E by weak stimulations arriving from two comparable initial delay routes; $\sim2\tau$ ms (A-F-E) and $\sim2\tau-\varepsilon$ ms (A-B-C-D-E) (**Figure 3A**). Initially, neuron E does not fire since $\varepsilon\approx1.7$ ms$>T_{TS}\approx0.5$ ms. Since the overall increase in the neuronal response latency of a chain is accumulative, proportional to the number of neurons it comprises, $\Delta(Stim_E)$ gradually decreases below $T_{TS}$ (**Figure 3B**) and neuron E suddenly starts to fire. Consequently, since neuron A fires every $\sim2\tau$ ms and neuron E fires $\sim2\tau$ ms laggard to A, $Sync_{AE}\approx0$ (**Figures 3B,C**). As $\Delta(Stim_E)$ decreases, the response of neuron E becomes more reliable (**Figures 3B,C**) and a leap out of synchrony is observed when $\Delta(Stim_E)$ again exceeds $\sim T_{TS}$ (**Figure 3B**). Since neuron E's firing does not close a new neuronal loop, the leaps in and out of synchrony do not affect the firing frequency of the neuronal circuit (**Figure 3C**). The number of spikes to synchrony increases with $\varepsilon$ as well as the time-gap between neurons during synchronization, $Sync_{AE}$ (**Figures 3D,E**). Simulation results (not shown) confirmed the robustness of the experimentally observed leap in and out of synchrony without a frequency change in population dynamics.

## DISCUSSION

Understanding the brain mechanisms that underlie firing synchrony is one of the great challenges of neuroscience. There are many variants of population codes, where a set of neurons in a population acts together to perform a specific computational task (Palm, 1990; Eichenbaum, 1993; Ainsworth et al., 2012). There is much discussion over whether *rate* coding or *temporal* coding is used to represent perceptual entities in populations of neurons in the cortex. A number of reports suggest that almost all the information in a stimulus is embedded in the rate code of active neurons (Aggelopoulos et al., 2005), while others suggest that synchrony among spiking of neuronal populations carry the information (deCharms and Merzenich, 1996). Experimental support for changes solely in firing rate when the perceptual task is modified (e.g., Lamme and Spekreijse, 1998; Roelfsema et al., 2004) is as compelling as those works that show changes in synchrony in the absence of firing rate changes (e.g., Womelsdorf et al., 2005), whereas in other experiments changes in both rate and spike correlations are observed concurrently (e.g., Biederlack et al., 2006). In any case, the usefulness of rate coding and temporal coding as information carriers of brain activity is a function of the decoding complexity, which is tightly correlated with their accuracy.

Rate and temporal coding are typically inaccurate in brain activities. Rate precision, measured by inter-spike interval (ISI) distributions, typically follows a broad distribution, deviating from a Poissonian one (Amarasingham et al., 2006). Similarly, relative spike timings between coactive neurons are inaccurate, typically within the precision of several milliseconds (Kayser et al., 2010; Wang, 2010). These types of inaccuracies indicate that the mission to grasp gradual changes in temporal and/or rate coding (e.g. changes from an average firing rate of 5 Hz to 6 Hz), on a timescale of a few ISIs, is a heavy computational mission which might not be satisfactorily resolved. The underlying cause of this computational difficulty is the broad distribution of the ISIs which is overlapped between gradually changed temporal codes or gradually changed rate codes.

To overcome this difficulty we proposed a mechanism which enables the emergence of a sudden leap to synchrony together with or independent of a leap in the firing frequency. This mechanism results in leaps from firing mismatches of several dozens of milliseconds to nearly zero-lag synchronization, and can be accompanied by a sudden frequency multiplication of the neuronal firing rate. These sudden changes occur on a time scale of extremely few ISIs, and are easily detectable as the distributions of the ISIs before and after the leaps are non-overlapping. Hence, one ISI is sufficient to detect the transition without accumulatively estimating the ISI distribution. These fast and robust indicators might be used as reliable information carriers of time-dependent brain activity.

The proposed mechanism also allows for the simultaneous emergence of sudden leaps in rate and temporal synchrony, hence bridging between these two major schools of thought in neuroscience (Eckhorn et al., 1988; Gray et al., 1989; Ahissar et al., 1992; Nicolelis et al., 1995). This mechanism requires *recurrent* neuronal circuits, and synchrony appears even among neurons which do not share a common drive. Sub-threshold stimulations (e.g. the stimulations to neuron C in **Figures 1,2** and to neuron E in **Figure 3**) serve as a switch that momentarily closes or opens a loop in the neuronal circuit. The state of the switch changes a global quantity of the network, the GCD of the entire circuit's loops, which determines the state of synchrony (e.g. zero-lag synchrony, cluster synchrony, shifted zero-lag synchrony) (Kanter et al., 2011; Nixon et al., 2012). These demonstrated prototypical examples call for a theoretical examination of more structured scenarios, including multiple leaps in and out of synchrony. In addition, a more realistic biological environment has to be examined containing synaptic noise and adaptation.

**APPENDIX**

## SLOW DIVERGENCE OUT OF SYNCHRONY

The slow increase in Sync$_{AB}$ (**Figure 1D**) is analytically examined below for a case of two phase-to-phase neurons, A and B, as depicted in **Figure A1.** The derivation below is in the spirit of Ermentrout's analysis of coupled type I membranes (Ermentrout, 1996). We first define the following quantities and assumptions:

$t_i(q) \equiv$ the timing of the $q^{th}$ spike of neuron i, e.g. $t_A(0)$ is the timing of the first spike of neuron A, where the count starts at 0.

$L_i(q) \equiv$ neuronal latency of neuron i at its $q^{th}$ spike.

The initial time delays are $\tau_{AB} = \tau_{BA} \equiv \tau$.

Assuming initial conditions, t=0, where both neurons fire simultaneously, i.e. $t_A(0) \equiv 0$, $t_B(0) = 0$.

The spiking times of neurons A and B are given by

$$\begin{cases} (i) & t_B(q) = t_A(q-1) + \tau + L_B(q) \\ (ii) & t_A(q) = t_B(q-1) + \tau + L_A(q) \end{cases}$$

Substituting (ii) into (i) and vice versa:

$$\begin{cases} t_B(q) = t_B(q-2) + \tau + L_A(q-1) + \tau + L_B(q) \\ t_A(q) = t_A(q-2) + \tau + L_B(q-1) + \tau + L_A(q) \end{cases}$$

one can find that the solution of these coupled recursive equations is given by:

$$\begin{cases} t_B(q) = \sum_{q'=1}^{\frac{q}{2}} L_B(2q') + L_A(2q'-1) + 2\tau \\ t_A(q) = \sum_{q'=1}^{\frac{q}{2}} L_A(2q') + L_B(2q'-1) + 2\tau \end{cases}$$

Consequently, the firing time-gap between the two neurons is given by

$$Sync_{AB}(q) \equiv |t_B(q) - t_A(q)|$$

$$Sync_{AB}(q) = \left| \sum_{q'=1}^{\frac{q}{2}} L_B(2q') + L_A(2q'-1) - L_A(2q') - L_B(2q'-1) \right|$$

$$Sync_{AB}(q) = \left| \sum_{q'=1}^{\frac{q}{2}} (L_A(2q'-1) - L_A(2q')) - \sum_{q'=1}^{\frac{q}{2}} (L_B(2q'-1) - L_B(2q')) \right|$$

Under the assumption of continuous increase in latency and large q

$$\frac{dL_i(2q')}{d(2q')} \approx -(L_i(2q'-1) - L_i(2q')) > 0$$

$$Sync_{AB}(q) = \left| \int_0^{\frac{q}{2}} \frac{dL_A(2q')}{d(2q')} \frac{dq'}{d(2q')} d(2q') - \int_0^{\frac{q}{2}} \frac{dL_B(2q')}{d(2q')} \frac{dq'}{d(2q')} d(2q') \right|$$

$$Sync_{AB}(q) = |0.5(L_A(q) - L_A(0)) - 0.5(L_B(q) - L_B(0))| \tag{A1}$$

Note that these calculations refer to even values of q. Similar equations can be obtained for odd values of q (not shown). In addition, fluctuations in the latencies may also enhance the deviation from synchronization.

## ACKNOWLEDGMENTS

We would like to thank Moshe Abeles and Evi Kopelowitz for stimulating discussions. Fruitful computational assistance by Yair Sahar and technical assistance by Hana Arnon are acknowledged. This research was supported by the Ministry of Science and Technology, Israel.

**Conflict of Interest Statement:** The authors declare that the research was conducted in the absence of any commercial or financial relationships that could be construed as a potential conflict of interest.

**FIGURE 1 | A sudden leap to synchrony accompanied by frequency doubling.** Notations used: $Sync_{AB}$, the absolute time-lag between the spikes of neurons A and B; $\Delta(Stim_C)$, the absolute time difference between two weak stimulations to neuron C; $\Delta L_D$, the increase in response latency of neuron D after n evoked spikes. **(A)** Schematic of a neuronal circuit consisting of four neurons and weak/strong stimulations represented by dashed (green)/full (black) lines. An initial stimulation is given to neuron A. **(B)** Experimental measurements of $\Delta(Stim_C)$ as a function of the spikes of neuron A. $\Delta(Stim_C)$ is initially set to $\varepsilon \approx 0.8$ ms (green line) with $\tau=50$ ms and $T_{TS} \approx 0.23$ (presented by the dashed horizontal green line). A unified longer stimulation was given in events where the time-lag between the weak stimulations<20μs (presented by $\Delta(Stim_C)=0$). $Sync_{AB}$ is presented by the blue line, indicating a sudden leap from $\tau=50$ ms to nearly zero-lag synchronization. **(C)** Spike trains of the four neurons. A sudden leap to $Sync_{AB}\approx 0$ occurs at time/$2\tau=122.5$ (at spike 121 of neuron A) immediately following a single evoked spike of neuron C. It is accompanied by a doubled firing frequency, from ~10 Hz to ~20 Hz. $Sync_{AB}\approx 0$ is robust to response failures of neuron C, e.g. time/$2\tau=124.5$. **(D)** $Sync_{AB}$ as a function of the spikes of neuron A, for various $\varepsilon$, where the data for $\varepsilon=0.8$ (blue) is the same as in **(B)** and **(C)**. The number of spikes to a leap to synchrony increases with $\varepsilon$. **(E)** $\Delta L_D$ for repeated stimulations at 10 Hz. $\Delta L_D$ at the synchrony leap for different $\varepsilon$ are colored following **(D)**. Note that $Spike_D$ is equal to $Spike_A$ in **(B),(D)**. **(F)** Results of population dynamic simulations where each neuron in **(A)** is now represented by a population comprised of 40 Hodgkin-Huxley neurons, each one innervated by 4 randomly chosen neurons from each of its driving clusters. The delays between neurons are taken from a Gaussian distribution centered at the delays of the single neuron case with a variance of 0.2 ms. For simplicity, each time a neuron fires all of its outgoing delays are increased by 0.04 ms. The simulation parameters were $\varepsilon=2$ ms and $T_{TS}\approx 1.3$ ms. **(G)** Raster plot of the 120 neurons comprising nodes A, B and C. A leap to synchrony occurs at time/$2\tau\approx 20$, accompanied by a doubling of the firing frequency.

**FIGURE 2 | A sudden leap to synchrony accompanied by tripled frequency.** Notations used: $Sync_{AB}$, the absolute time-lag between the spikes of neurons A and B; $\Delta(Stim_C)$, the absolute time difference between two weak stimulations to neuron C; $\Delta L_D$, the increase in response latency of neuron D after n evoked spikes. **(A)** Schematic of a neuronal circuit as in **Figure 1A**, however the delay from neuron B to A is now $2\tau$. **(B)** Experimental measurements of $\Delta(Stim_C)$, similar to **Figure 1B**, with $\varepsilon\approx 0.5$ ms, $3\tau=125$ ms and $T_{TS}\approx 0.2$ (presented by the dashed horizontal green line). $Sync_{AB}$, (blue line) indicating a sudden leap from $\tau\approx 125/3$ ms to nearly zero-lag synchronization. **(C)** Spike trains of the four neurons. A sudden leap to synchronization, $Sync_{AB}\approx 0$, occurs at time/$3\tau=44$ (at spike 44 of neuron A) consecutive to three evoked spikes of neuron C. This is accompanied by tripled firing frequency of neurons A and B, from ~8 Hz to ~24 Hz. $Sync_{AB}\approx 0$ is robust to response failures of neuron C, e.g. time/$3\tau=46.33$. **(D)** $Sync_{AB}$ as a

function of the spikes of neuron A for various ε, where the number of spikes to the leap to synchrony increases with ε. The data for ε=0.5 (blue) is the same as in **(B)** and **(C)**. The observed oscillations in $Sync_{AB}$ before a leap to synchrony originate from response failures of neuron C, and similarly oscillations in a leap out of synchrony originate from response failure of either neuron A or B. **(E)** $\Delta L_D$, for repeated stimulations at 8 Hz. $\Delta L_D$ at the leap for different ε are indicated and colored following **(D)**, approximately verifying Equation 1, e.g. for ε=0.8 ms and $T_{TS}$≈0.2 ms, $\Delta L_D(197)$ gives ~0.6 ms. Note that $Spike_D$ is equal to $Spike_A$ in **(B)**,**(D)**. **(F)** Results of population dynamic simulations similar to **Figure 1F,G** with ε=2 ms, $T_{TS}$≈1.3 ms and 3τ=125 ms. **(G)** Raster plot of the 120 neurons comprising nodes A, B and C. A leap to synchrony occurs at time/3τ≈20, accompanied by tripled firing frequency.

**FIGURE 3 | Short epochs of synchrony not accompanied by a change in frequency.** Notations used: $Sync_{AE}$, the absolute time-lag between the spikes of neurons A and E; $\Delta(Stim_E)$, the absolute time difference between two weak stimulations to neuron E; ΔL, defined as $\Delta L_B + \Delta L_C + \Delta L_D - \Delta L_F$. **(A)** Schematic of a neuronal circuit consisting of six neurons and weak/strong stimulations represented by dashed (green)/full (black) lines. **(B)** Experimental measurements of $\Delta(Stim_E)$, similar to $\Delta(Stim_C)$ in **Figure 1B**, with ε≈1.7 ms, τ=50 ms and $T_{TS}$≈0.5 ms (presented by the dashed horizontal green line). The time delay between neurons A and E, ~2τ, is denoted by the dashed horizontal black line. The firing region of neuron E (blue dots bounded by dashed vertical guidelines), which is at nearly zero-lag synchronization with the firing of neuron A, $Sync_{AE}$≈0, starts after 77 spikes of neuron A. The temporary firing of E terminates after ~200 spikes of neuron A. **(C)** Spike trains of neurons A, F and E, indicating a steady firing frequency (~10 Hz) of the neuronal circuit independent of the firing of neuron E, where an epoch of synchrony, $Sync_{AE}$≈0, begins at time/2τ=77 (at spike77 of neuron A). **(D)** The number of spikes prior to the firing of neuron E increases with ε. The mild increase in the firing mismatch, $Sync_{AE}$, is attributed to the additional increase by ε of the initial 2τ delay loop (E fires ~2τ+ε laggard to A, however the time-gap between consecutive firings of A is ~2τ+2ε). The data for ε=1.7 (blue) is the same as in **(B)** and **(C)**. **(E)** ΔL for repeated stimulations at 10 Hz. ΔL at the synchrony leap for different ε are colored following **(D)**. The number of spikes per neuron (e.g. $Spike_A$), n, until the leap to synchrony increases with ε and can be obtained from Equation 1, where $\Delta L_D$ is substituted by ΔL.

**FIGURE A1 | Slow divergence out of synchrony between two phase-to-phase neurons** Notation used: $Sync_{AB}$, the time-lag between the spikes of neurons A and B. **(A)** Schematic of two bidirectional interconnected spiking neurons. The initial delays between the neurons are equal, $\tau_{AB}=\tau_{BA}=\tau$. **(B)** Response latency of both neurons as a function of spike number. The latencies were taken to be $L_A=0.5*\ln(q+5)+2$, $L_B=0.3*\sqrt{q+2}+3$, qualitatively similar to latency profiles observed in experiments. **(C)** $Sync_{AB}$ as a function of spike number for the latencies depicted in **(B)**, assuming $Sync_{AB}(0)=0$. The calculation (brown line) was done using Equation A1 and is in a good agreement with straightforward simulations of exact spike times (black dots). For simplicity, the simulated $Sync_{AB}$ is only displayed for even numbers of spikes.

FIGURE 1

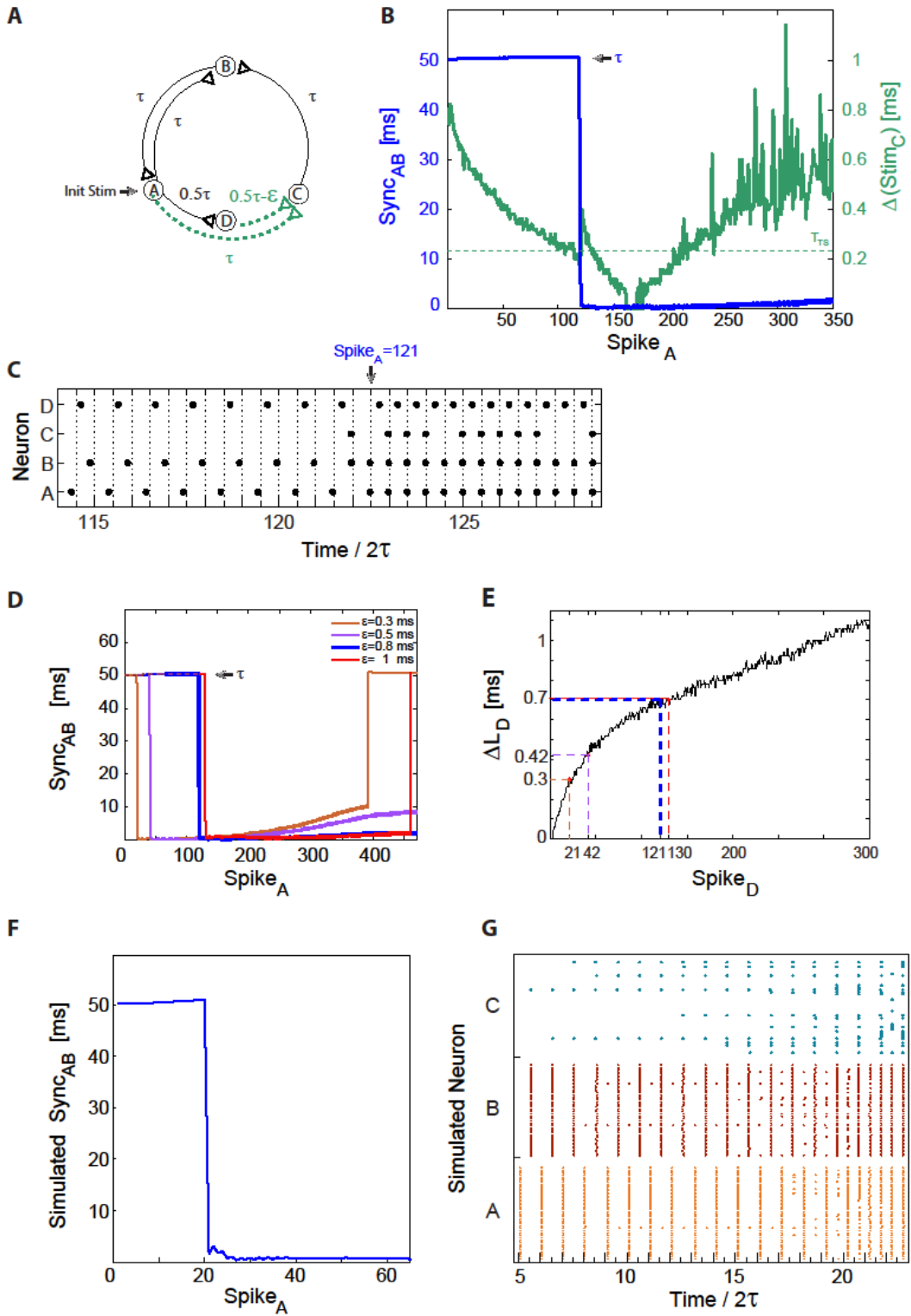

FIGURE 2

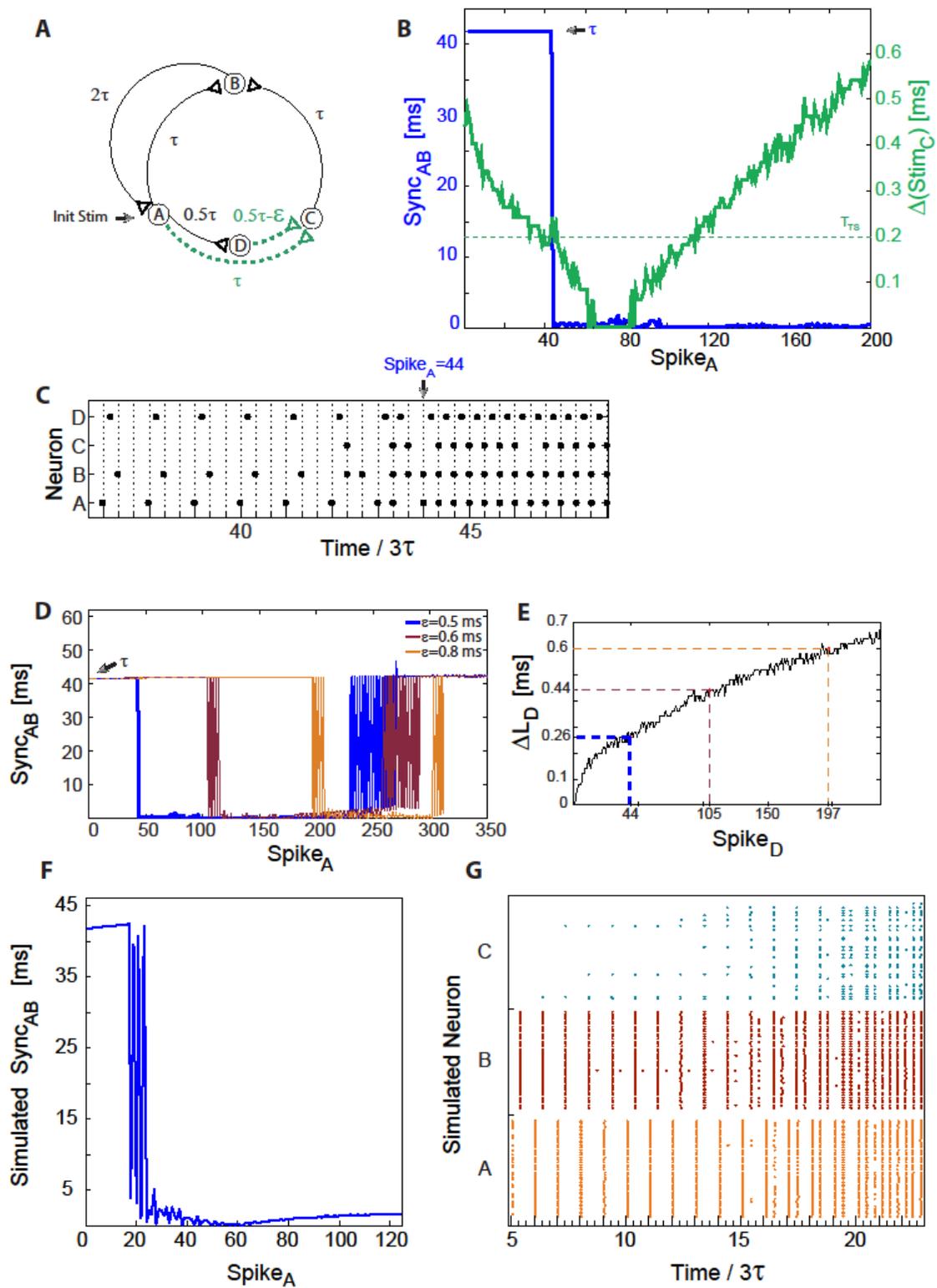
21

FIGURE 3

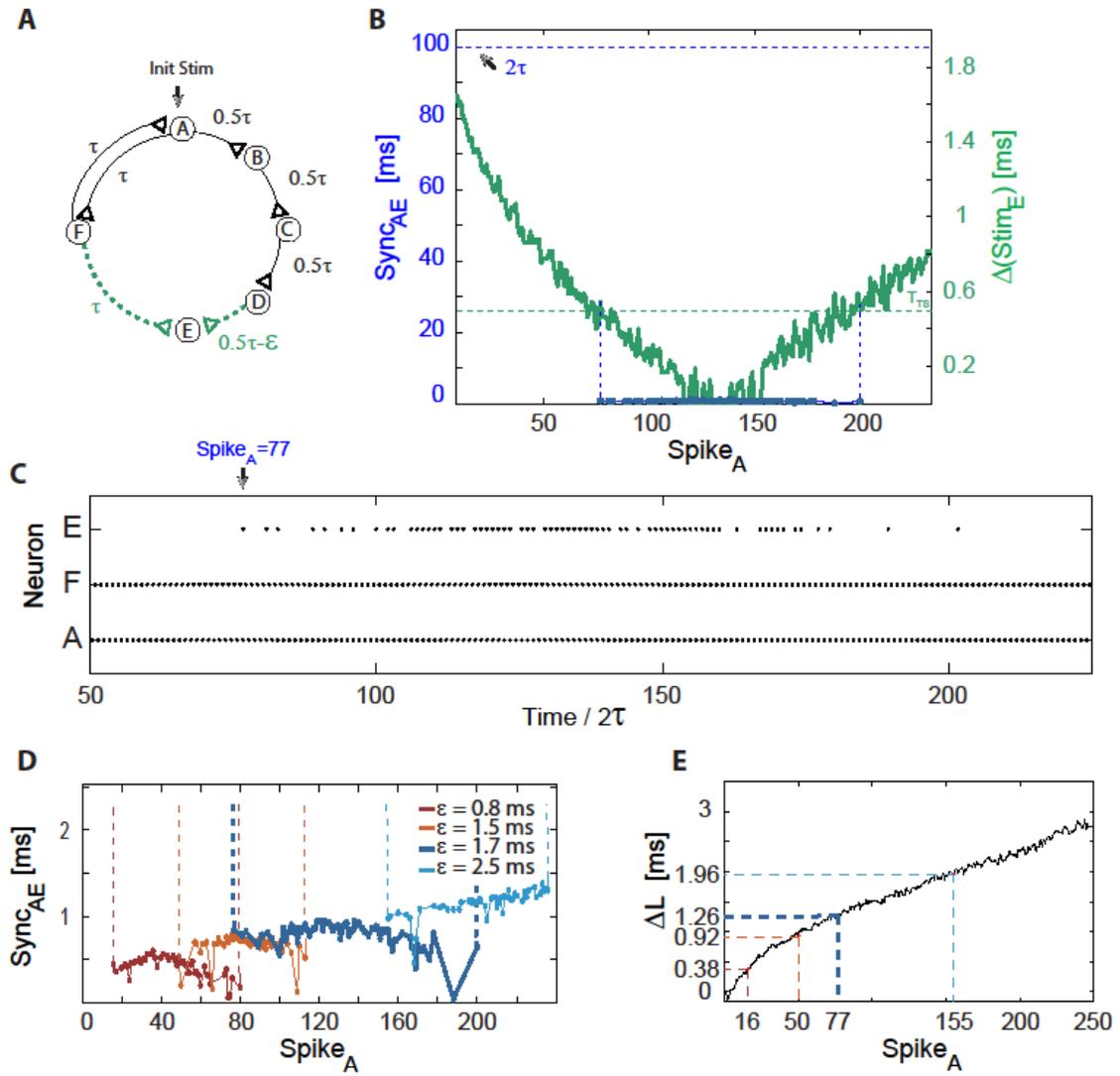